# Decrypting material performance by wide-field femtosecond interferometric imaging of energy carrier evolution

Pin-Tian Lyu, Qing-Yue Li, Pei Wu, Chao Sun, Bin Kang*, Hong-Yuan Chen and Jing-Juan Xu*

State Key Laboratory of Analytical Chemistry for Life Science, School of Chemistry and Chemical Engineering, Nanjing University, Nanjing 210023, China

*Bin Kang and Jing-Juan Xu

**Email:**  binkang@nju.edu.cn (B.K.); xujj@nju.edu.cn (J.-J.X.)

**ABSTRACT**

Energy carrier evolution is crucial for material performance. Ultrafast microscopy has been widely applied to visualize the spatiotemporal evolution of energy carriers. However, direct imaging of small amounts of energy carriers on nanoscale remains difficult due to extremely weak transient signals. Here we present a method for ultrasensitive and high-throughput imaging of energy carrier evolution in space and time. This method combines femtosecond pump-probe techniques with interferometric scattering microscopy (iSCAT), named Femto-iSCAT. The interferometric principle and unique spatially-modulated contrast enhancement increase the transient image contrast by >2 orders of magnitude and enable the exploration of new science. We address three important and challenging problems: transport of different energy carriers at various interfaces, heterogeneous hot electron distribution and relaxation in single plasmonic resonators, and distinct structure-dependent edge state dynamics of carriers and excitons in optoelectronic semiconductors. Femto-iSCAT holds great potential as a universal tool for ultrasensitive imaging of energy carrier evolution in space and time.

**INTRODUCTION**

Energy carrier evolution is essential to catalysis, sensing and energy conversion. In general, the transport, distribution and relaxation of energy carriers directly influence the material functionality. In recent years, ultrafast optical microscopy combines pump-probe spectroscopy with microscopy techniques, and has emerged as a powerful tool for imaging spatially-dependent dynamics and tracking the transport of excited-state charge carriers and excitons.[1–4] However, detection of weak transient signals has always been challenging because of the small sample interaction volume.[5,6] For instance, transient absorption microscopy (TAM) based on either beam or sample scanning requires lock-in detection to extract weak signals from noise.[7] An alternative approach for TAM is to use wide-field illumination with a high-speed detector.[8] Although wide-field TAM benefits from simple setups and short acquisition time, it usually yields lower signal-to-noise ratio (SNR). Increasing the pump intensity could improve the signal level by generating more excited-state carriers, but it may also lead to multi-photon effects or thermal damage. For wide-field ultrafast microscopy, the transient signal is directly extracted from the images pairs with the pump beam on and off. In other words, the differential image contrast (pump induced change in probe intensity, $\Delta I/I$) contains the whole spatiotemporal information of the signal. Hence, contrast enhancement opens a possible path for transient signal amplification.

Interferometry is one of the most effective methods for weak signal detection in many fields, ranging from gravitational wave detection,[9] nano-displacement measurement[10] to ultrasensitive optical imaging.[11] In particular, interferometric scattering microscopy (iSCAT) has taken advantage of the interference between the object-scattered light and the reference light to achieve sufficient signal contrast of single nanoparticles[12] and even single molecules.[13] Demonstrating iSCAT's feasibility in time-resolved imaging, stroboscopic scattering microscopy (stroboSCAT)[14] has been developed with the image contrast around $10^{-4}$ and the temporal resolution of ~240 ps for tracking of energy flow in semiconductors. However, there is still much room for the breakthrough in the image contrast as well as the temporal resolution. In well-designed stroboSCAT experiments, shot noise is the dominant noise source. Thus, the SNR can be improved by increasing the number of detected signal photons. One approach is to integrate thousands of frames into one image as stroboSCAT has done. The other approach is by increasing the intensity of the probe beam so that the scattering signal increases. However, in common-path geometry, the intensity of the reference beam also scales with the probe intensity. Consequently, to avoid saturating the camera, the achievable sensitivity is limited by the maximal reference beam intensity that the camera can take. To further improve the sensitivity, one can selectively reduce the intensity of the reference beam without attenuating the scattering signal by inserting a spatial filter (e.g., a dot-shaped partial reflector[15–17]) so that the sample can be illuminated at a higher probe intensity. In this case, the optical noise is kept with no change while the SNR is increased. The samples are tolerant of the increase in probe intensity because we detect the scattering signal rather than resonant absorption. Thus, in principle, the image contrast in ultrafast iSCAT could be further enhanced by spatially modulating the interferometric field accompanied by increasing the probe intensity, allowing ultrasensitive detection of a small amount of energy carriers.

Here we report wide-field femtosecond interferometric scattering microscopy (Femto-iSCAT) as a universal, ultrasensitive and high-throughput imaging method to visualize the transport, distribution and relaxation of energy carriers. This is the first time that iSCAT has been combined with femtosecond pump-probe techniques to realize interferometric imaging of energy carriers in the femtosecond realm. Without increasing the pump intensity, the transient image contrast is enhanced by >2 orders of magnitude compared to previous methods via interference with spatial modulation and thus help reveal new scientific phenomena. We showcase extensive and challenging applications of Femto-iSCAT, including concurrent tracking of different energy carriers (charge, heat, etc.) at optoelectronic interfaces, mapping of hot electron heterogeneity in single plasmonic resonators, and imaging of carrier and exciton decay through distinct edge states in new semiconductors. Femto-iSCAT is promising as a universal and ultrasensitive tool for study of various ultrafast processes.

## RESULTS AND DISCUSSION

**Principle of Femto-iSCAT.** As shown in Figure 1a, the detected signal in iSCAT is generated by the interference between the light scattered by the sample ($E_s$) and the reference light from the interface ($E_r$). Inspired by the principle of iSCAT, Femto-iSCAT (Figure 1b) uses a femtosecond laser as the coherent light source. We first excite the sample with a focused or wide-field pump pulse to generate energy carriers, which act as point scatterers and change the local dielectric constants ($\Delta\varepsilon$). Subsequently, we use a wide-field probe pulse to detect the changes to the scattering amplitude ($\Delta|s|$) of the excited area at a series of time delays, thus imaging the spatiotemporal evolution of energy carriers. The probe (central wavelength at 730 nm) is spectrally away from the material band edge to primarily detect the change in $\varepsilon$, avoiding resonant absorption. The point spread function (PSF) manifests as a typical interferometric fringe and fits the 2D Bessel function. This feature distinguishes our method from the reflection or transmission based imaging such as laser scanning confocal TAM, where the PSF is a Gaussian function.

We show the simplified setup configuration of Femto-iSCAT in Figure 1c. Implementations of two pump modes fulfil the requirements of various applications. In the focused pump mode, a diffraction-limited pump is used to excite the sample so that time-evolving transport of local energy carriers is generally concerned. In the wide-field pump mode, a large area is pumped so that the whole scattering profile is changed. According to Huygens' principle, wide-field pump-induced changes to the scattered field originate from the superposition of all wavelets from point scattering sources. As a result, the spatial distribution of energy carriers is visualized. Both modes are equipped with a partial reflector (PR) as a spatial filter to enhance the image contrast. As shown in Figure 1d, the reference field is selectively attenuated by the partially reflective gold layer, while the scattered field is mostly transmitted to the image plane. It is known that in contrast-enhanced iSCAT, the transmissivity ($t$) of the PR determines its enhancement factor. As a proof of concept, we used the PR with $t$ = 10% (several to ten fold enhancement as previously shown in iSCAT) in consideration of the optimal photon flux at the camera. It has achieved enough sensitivity to image hot carrier distribution in monolayer graphene that is usually difficult to visualize due to the extremely low signal, only with an acquisition time less than 30 seconds. If necessary, the contrast could be further enhanced by rational design of the PR together with a faster camera[15].

In comparison to femtosecond transient absorption microscopy, Femto-iSCAT is sensitive to the change to the real part of the refractive index ($n$) instead of the imaginary part of the refractive index ($k$, which is related to absorption). Thus, Femto-iSCAT is more generalizable to diverse materials (especially for thick or opaque samples) and energy forms. Additionally, the probe need not to be set to a specific resonant absorption wavelength, which reduces the thermal effect and also makes it feasible to obtain in situ spatial and spectral information concurrently. Moreover, Femto-iSCAT has several unique features. First, interferometric enhancement enables ultrasensitive detection of weak signals with short acquisition time. Second, wide-field imaging captures a snapshot of a large area, which proves to be a high-speed and high-throughput method. Finally, the phase sensitivity of interferometric scattering is expected to realize 3D imaging of energy carriers.

**Energy carrier transport at interfaces.** Energy carrier transport at the material surface and interface plays an essential role in photocatalysis and photovoltaic conversion. Direct imaging of energy carrier transport has been achieved by ultrafast electron microscopy,[18] time-resolved optical microscopy,[19] time-resolved photoemission electron microscopy,[20] etc. However, it is still challenging to detect transient signals of samples with tiny absorption cross section or low damage threshold. Here we utilized Femto-iSCAT under the focused pump mode to track interfacial energy carrier transport with high sensitivity and sample universality. The Femto-iSCAT images manifest as interferometric fringes due to the enhanced interferometric field. To simplify the analysis, only the central fringes are used to investigate the spatiotemporal dynamics. The signal profiles exhibit a temporally decaying 2D Gaussian distribution convolved with the PSF of the system. However, the latter is invariant over time, so it would not affect the result of the diffusion constant, $D$, which is determined by the variance $\sigma^2$ of the Gaussian profile along one dimension with time $t$: $2Dt = \sigma^2(t) - \sigma^2(0)$ for isotropic diffusion.[21]

Hot electron diffusion dynamics in plasmonic nanomaterials are of great importance for energy transfer and heat management in nanodevices.[22,23] Direct imaging of hot electron diffusion on metal surfaces has only been achieved by scanning ultrafast microscopy based on transient reflection.[24,25] In those works, a high pump fluence of 1 mJ cm$^{-2}$ was used for the gold films (thickness>50 nm) to ensure the contrast up to 10$^{-4}$. For ultrathin films (e.g., thickness<10 nm), low pump fluence is necessary to avoid thermal damage, however, predictably leading to a much weaker signal. Here we explored the diffusion dynamics of hot electrons and thermal energy in a 10-nm thin gold film (Figure 2a). The interaction between electron-phonon (e-ph) coupling and hot electron diffusion leads to electron relaxation and thermal equilibrium within several picoseconds (Figure 2b). The image contrasts in the initial Femto-iSCAT images (Figure 2c) reach the order of 10$^{-2}$, increasing by 2 orders of magnitude at a 10-fold lower pump fluence (0.1 mJ cm$^{-2}$) compared to the transient reflection method at similar noise limit. The relaxation dynamics show an exponential decay with $\tau$ = 1.9 ps, which denotes the electron-lattice thermalization time (Figure 2d). To uncover the spatial diffusion behavior, we fitted the signal line profiles with Gaussian functions (Figure 2e). We extracted the Gaussian distribution variance $\sigma^2$ and plotted the time evolution of $\sigma^2(t) - \sigma^2(0)$ as a function of time delay (Figure 2f). The results show an initial fast diffusion followed by a slower spreading after ~5 ps. The diffusion constants $D_{fast}$ and $D_{slow}$ were determined to be 109 cm$^2$ s$^{-1}$ and 1.2 cm$^2$ s$^{-1}$, respectively. These two processes are attributed to hot electron diffusion and phonon-limited thermal diffusion, respectively. Our results are much closer to the theoretical electron and thermal diffusivities compared to the previous report [24]. We could also derive the electronic thermal conductivity of gold, $k$, from $D_{slow}$ (known as the thermal diffusivity of gold) and $C_l$ (the lattice heat capacity of gold): $k \approx D_{slow}C_l$ = 294 W m$^{-1}$ K$^{-1}$, which is close to the acknowledged value [24]. We show that Femto-iSCAT enables ultrasensitive spatiotemporal tracking of hot electron and heat diffusion in ultrathin metal nanomaterials, which facilitates the understanding of electron-phonon interactions and heat dissipation in nanomaterials and thereby inspires the design of thermoelectric and optoelectronic nanodevices.

The migration of charge carriers in silicon-based materials is fundamental to modern electronics, optoelectronics and solar energy conversion.[26,27] Carrier diffusion in silicon has been investigated by surface photovoltage,[28] scanning ultrafast electron microscopy,[29] etc. However, these techniques mainly focus on only a few carrier types and rely on specific optoelectronic properties of samples. Here we used Femto-iSCAT to visualize charge carriers and heat transport in n-doped silicon. Photoexcitation creates a neutral carrier cloud consisting of free electrons and holes, which spread rapidly. The interplay between carrier diffusion and recombination determines carrier dynamics (Figure 2g). In silicon, electron-hole pairs mostly recombine through defects,[30] causing lattice heating (Figure 2h). The Femto-iSCAT time series (Figure 2i) show distinct diffusion features at a pump fluence of 0.2 mJ cm$^{-2}$: a negative signal expands fast at first, and a positive signal appears later in the center, indicating that different species are generated. The results enrich the details of carrier evolution within femtosecond to picosecond time scales beyond the temporal resolution limit of stroboSCAT.[14] In addition, the image contrasts are about 2 orders of magnitude higher than those in stroboSCAT with half the pump fluence (0.2 mJ cm$^{-2}$ in Femto-iSCAT). The transient signal decays with biexponential functions, $\Delta I/I = A_1 e^{-t/\tau_1} + A_2 e^{-t/\tau_2}$ (Figure 2j). The average lifetime of these two fast components was calculated as $\tau_{Avg} = (A_1\tau_1 + A_2\tau_2)/(A_1 + A_2)$. This decay ($\tau_{Avg}$ = 110 ps) is attributed to free carrier recombination and diffusion. We observed a clear transition of the signal sign in the signal profiles (Figure 2k). The diffusion constants can be extracted from the time evolution of $\sigma^2(t) - \sigma^2(0)$ (Figure 2l). The negative and positive signals expand at rates of 30 cm$^2$ s$^{-1}$ ($D_{fast}$) and 0.6 cm$^2$ s$^{-1}$ ($D_{slow}$), respectively. According to the reported values,[14] we attribute these two signals to free electrons and heat deposited in the lattice, which are supposed to produce opposite changes to the dielectric constant of silicon. We note that this contrast reversal is not caused by axial diffusion because axial diffusion at the nanoscale would continuously tune the contrast, which was not observed here. These diffusion parameters are important for the rational design of silicon-based optoelectronic devices. We therefore prove Femto-iSCAT's superiority in tracking carrier and heat transport in semiconductors with high sensitivity as well as high temporal resolution.

Perovskites have attracted considerable attention as promising optoelectronic materials due to their long carrier lifetime and extended carrier diffusion lengths.[31,32] Particularly, carrier diffusivity is critical for the development of novel perovskite solar cells.[33] Transient absorption microscopy has been widely used to track carrier transport in perovskites,[34] however, pump intensity has to be well controlled because perovskites are vulnerable to photodamage. Here we used Femto-iSCAT to track long-lived carrier transport in fragile photovoltaic materials. An optical image exhibits an optically flat CsPbBr$_3$ microcrystal on glass substrate (Figure 2m). Carrier transport in the CsPbBr$_3$ microcrystal was visualized by Femto-iSCAT (Figure 2o). The image contrasts reach the order of 10$^{-2}$ at a very low pump fluence of 10 μJ cm$^{-2}$. Excited-state carriers primarily recombine through photoluminescence (PL) in CsPbBr$_3$ (Figure 2n). The slow decay ($\tau_{free\ carrier}$ = 1236 ps) is attributed to free carrier recombination (Figure 2p). The fast relaxation (<200 ps) is probably related to trap states or nondiffusible excited-state species,[35] which is not of our concern. The line profiles of the time-evolving carrier density are shown in Figure 2q. We determined the diffusion constant to be 0.2 cm$^2$ s$^{-1}$ (Figure 2r). The linear dependence reveals free carrier diffusion behavior. In this case, the carrier density exceeds the trap density so that all traps are filled in the first few nanoseconds. The diffusion constant is lower than that in perfect

CsPbBr$_3$ single crystals (1 cm$^2$ s$^{-1}$)[14] because defects and grain boundaries hinder carrier diffusion. Femto-iSCAT shows great potential for tracking energy carrier transport in perovskites with low damage threshold, which benefits the design of next-generation photovoltaic devices.

**Heterogeneous hot electron dynamics.** The physical and chemical properties of nanomaterials are closely related to the dynamics of charge carriers and phonons.[36,37] The size, shape and structure of nanoparticles are inhomogeneous in the ensemble. Measuring the distribution and relaxation of charge carriers and phonons in single nanoparticles offers insight into the structure-property relationships. Although a single nanoparticle gives an extremely weak transient signal due to its tiny optical cross section,[38] Femto-iSCAT, benefiting from high sensitivity and temporal resolution, potentially provides new perspectives for single-particle dynamics. The high-throughput wide-field pump mode helps correlate dynamical behaviors with specific structural features, shedding light on the heterogeneity of individual particles in the ensemble, even at different locations within a single particle.

Hot electrons in a plasmonic nanoparticle decay through electron-phonon coupling, which leads to the coherent excitation of vibrational modes of the particle. The frequencies of the vibrational modes depend on the size, shape and elastic constant of the particle.[39] Here we studied the ultrafast dynamics of single 20 nm gold nanospheres (Figure 3a). Figure 3e displays an iSCAT image of several dispersed 20 nm gold nanospheres on a glass coverslip. The Femto-iSCAT image of these particles, i.e., the hot electron map, is shown in Figure 3i. Note that only a few hundreds of image pairs (acquisition time <1 min) were averaged to achieve sufficient contrast. The visible variations in hot electron density indicate the heterogeneity of individual particles. A kinetic trace of a single particle is presented in Figure 3m. The initial fast decay (within a few ps) is ascribed to electron-phonon coupling, followed by damped acoustic vibration. The oscillating trace was fitted with an exponential decay superposed on a damped oscillation term:

$$\Delta I/I = A_c \exp(-t/\tau_c) + A_{br} \cos(2\pi t/T_{br} - \phi_{br}) \exp(t/\tau_{br}) \qquad (1)$$

where the first term accounts for the cooling of the particle from electron-phonon coupling and phonon-phonon coupling, and the second term represents the damped breathing mode, which gives a period of $T_{br}$ = 7.2 ps. This value is only slightly higher than the theoretical vibrational period (~6.6 ps) of 20 nm gold nanospheres.[39] We attribute the deviation to the size variation of individual particles. Owing to the high sensitivity of Femto-iSCAT, ultrafast dynamics of a single nanoparticle with such a small size are no longer difficult to visualize. Unravelling the electronic and elastic properties of single plasmonic nanoparticles will help guide the design of new optomechanical nanodevices.

Hot electron injection is one of the possible mechanisms of surface plasmon enhanced catalysis.[40,41] A spatially heterogeneous distribution of hot electrons in a plasmonic catalyst may lead to the spatial heterogeneity of catalytic enhancement, forming catalytic hotspots.[42,43] The plasmonic resonance mode has been extensively explored by several different methods, such as scanning near-field optical microscopy,[44] electron energy-loss spectroscopy,[45] two-photon luminescence microscopy,[46] cathodoluminescence nanoscopy,[47] etc. However, these methods usually only provide the map of hot electrons but not the relaxation lifetime. Herein, Femto-iSCAT is able to predict the spatial patterns of catalytic hotspots on single plasmonic particles by imaging hot electron distribution and the relaxation lifetime. The single gold nanowires, triangular and hexagonal gold nanoplates (Figure 3b-d) manifest special patterns in iSCAT images (Figure 3f-h) because of standing wave modes of plasmon resonance in microresonators.[48] The standing wave modes require the illumination by a coherent light source, otherwise the images show the reflection from the particle surfaces. In principle, standing plasmonic charge density waves confine optical energy and carriers in a localized region and form hotspots of electrons. The correlated hot electron maps obtained by Femto-iSCAT show that hot electrons concentrate at particular regions in these resonators (Figure 3j-l). We note that the signals outside the sample are the interferometric fringes as the rings in Figure 2. To uncover the spatial features of hot electron relaxation, we mapped the distributions of hot electron lifetime (Figure 3n-p). Because the amplitude of interferometric fringes scales with the signal of the particles, the lifetime maps show the associated fringes outside the particles. These fringes would not affect the extraction of spatially resolved dynamics, which is common for single-microparticle[49] and single-cell[50] analysis in iSCAT. In this gold nanowire, hot electrons concentrate at both ends with slightly longer lifetimes. Strikingly, in both triangular and hexagonal gold nanoplates, hot electron densities are higher at specific regions where hot electron lifetimes are prolonged to ~4 ps. This value is much longer than the hot electron lifetime (1~2 ps) in the other regions, which is in line with the typical lifetime in gold nanoparticles.[51] The special patterns of hot electron density are consistent with the distributions of the steady-state in-plane local density of states (LDOS) in these resonators, which depends on their physical size and geometric boundary.[48] According to the two-temperature model,[51] the electron-phonon coupling time depends on the initial electron temperature. Thus, hot electron lifetimes are prolonged at the regions with higher electronic energy. The locations with higher hot electron density and longer hot electron lifetime hold the potential to be the catalytic hotspots. The heterogeneous hot electron distributions and lifetimes in 1D and 2D plasmonic microresonators are observed for the first time. Femto-iSCAT provides direct evidence for characterizing the hot electron density and lifetime in plasmonic hotspots. The results show promise in revealing the underlying mechanisms of cavity physics, plasmonic catalysis and hot-electron chemistry.

**Edge state dynamics in semiconductors.** Lead halide perovskite crystals and 2D transition metal dichalcogenides (TMDCs) have attracted much interest due to their unique and extraordinary optoelectronic properties related to exciton and carrier behaviors. The diffusion, dissociation and recombination of excitons and carriers play an indispensable role in many applications, such as photovoltaic cells[52] and light-emitting devices.[53] It is well known that the edge states in semiconductors exhibit unique optical and electrical features. On the one hand, structural defects at the crystal edges form trap states[54] and thus restrict the mobility of free carriers, which may be detrimental to photovoltaic efficiency. On the other hand, at the few-layer edges of 2D materials,[55] additional electronic states are introduced to the band structure because of the presence of dangling bonds, usually increasing the carrier lifetime and improving the optoelectronic performance. However, little is known about ultrafast edge-state exciton and carrier dynamics in emerging semiconductors so far. Wide-field Femto-iSCAT is able to resolve subtle spatial difference of energy carrier density and lifetime at material surface. Here, for the first time, we provide direct evidence for the distinct pathways of energy carrier relaxation at different edge states in $CsPbBr_3$ microcrystals and 2D $WS_2$ layers.

The outline of a single $CsPbBr_3$ microcrystal is surrounded by the white dashed line in the optical image (Figure 4a). The Femto-iSCAT images show interesting dynamical behaviors of charge carriers at the crystal edges (Figure 4b). The edges exhibit significant signals far before the time zero (-8 ps), indicating that there exists particular emission process. Because the band-edge emission (~2.34 eV) is mostly spectrally filtered in the optical system and lasing would not occur under the low pump fluence regime, the subgap trap-state emission may dominate the signals at the crystal edges.

Then we gain insight into how the trap states affect the ultrafast carrier dynamics. The emission signals are invariant over the time range in the experiment, thus they are considered as a constant background to the transient signals. The carrier distribution is inhomogeneous on the crystal surface due to the surface defects that are invisible in the optical image. We extracted the kinetic traces from the edge and a smooth region at the interior (Figure 4c). Within the tens to hundreds of picosecond scale, the ultrafast processes are mainly attributed to nonradiative processes such as surface trapping[56] or carrier-phonon scattering,[57] while Auger recombination is negligible at low carrier density (<$10^{18}$ cm$^{-3}$). Notably, the carriers at the edges have a much shorter lifetime probably due to abundant trap states formed by structural defects. These shallow subgap traps slow down the carrier mobility by repeatedly capturing and thermally releasing the carriers,[58] which leads to fast decay of the dielectric constants, hence the transient signals. In contrast, the carriers at the interior are less affected by the surface traps and decay slower because there are less crystalline imperfections in the smooth facet. Note that this phenomenon varies from particle to particle. The spatial difference of carrier dynamics would be less conspicuous in relatively perfect crystals. We summarized the major recombination mechanisms in Figure 4d. The nonradiative relaxation of the carriers competes with the radiative recombination processes, but does not quench them completely, demonstrating that $CsPbBr_3$ crystals are highly tolerant of structural defects and traps.[59] Such feature helps them maintain good optoelectronic quality. Our results can provide insight into the relationship between carrier dynamics and surface defects in space and time, and also provide guidance for material synthesis to reduce or passivate the defects at the crystal edges.

The results in 2D materials tell a different story. The optical image of an exfoliated $WS_2$ crystal is shown in Figure 4e. The monolayer (1L), bilayer (2L) and trilayer (3L) $WS_2$ are surrounded by white dashed lines. Strongly bound excitons are considered as the primary excited-state species in 2D TMDCs because of quantum confinement and reduced dielectric screening. The Femto-iSCAT images show that the exciton density increases with the layer number, which is caused by larger absorption cross sections in thicker flakes (Figure 4f). Particularly, in 2L and 3L $WS_2$, the exciton density is significantly higher at the layer edges, indicating that lower-energy states probably exist at the few-layer edges. Structural defects are not associated with these states because the signal is homogeneous in 1L $WS_2$.

To unravel the mechanism of heterogeneous exciton distribution, we fitted the kinetic traces extracted from different positions with biexponential functions (Figure 4g). Generally, exciton-exciton annihilation (Auger recombination) dominate the fast decay at high excitation intensity.[60,61] The short lifetime $τ$ is considered as the nonradiative exciton lifetime.[62] The subsequent slow decay is ascribed to the thermal effect[63] or trapped carrier recombination,[64] neither of which interferes with the exciton lifetime. It is found that $τ$ is longer in thicker flakes. The thickness dependence can be explained as follows.[65] First, many-body interactions are stronger at the monolayer limit. Second, exciton-exciton annihilation requires phonon assistance in 2L and 3L $WS_2$. In other words, exciton-exciton annihilation is a less probable event in thicker flakes.

Moreover, we discovered that $τ$ is longer at the layer edges than that in the interior in 2L and 3L $WS_2$, but it is almost identical at different regions in 1L $WS_2$ (Figure 4g). Edge states in 2D layers for n>1 (Figure 4h) may contribute to the distinct properties at few-layer edges. We speculate that the key mechanism is trapping of excitons to the lower-energy edge states located at the layer edges, as shown in Figure 4i. Excitons can either decay through normal pathways (dominant for n=1) or be trapped at layer edge states (dominant for n>1). Once excitons are trapped at layer edge states, they are protected from losing energy through nonradiative recombination, which leads to a longer exciton lifetime. Thus, we uncover the spatial heterogeneity of exciton dynamics in layered 2D $WS_2$ using Femto-iSCAT and present a possible mechanism involving layer edge states, which may facilitate the understanding of optoelectronic properties at the edges and surfaces in 2D materials.

**CONCLUSIONS**

In summary, we presented a universal, ultrasensitive and high-throughput ultrafast microscopic technique, Femto-iSCAT, for imaging energy carrier evolution in space and time. Compared to previously reported stroboSCAT,[14] we primarily achieved three technical breakthroughs. First, we confirmed the viability of extending iSCAT to the femtosecond realm. Temporal resolution of 660 fs (360-fold better) enables measurements of faster processes such as hot carrier relaxation and nonradiative recombination. Second, we introduced an effective contrast enhancement approach based on spatial filtering. At several times lower pump intensity, the transient image contrast reaches the order of $10^{-2}$, which increases by >2 orders of magnitude. In addition, Femto-iSCAT provides both focused and wide-field pump modes to realize different functions. As a useful complement to tracking of diffusive processes, wide-field mapping of spatially-dependent dynamics is demonstrated to be informative and high-throughput for ultrafast imaging. These significant technical advances qualify us for observation and measurements of various novel photo-induced processes.

We show a variety of challenging applications to prove the advance of Femto-iSCAT. We first leveraged Femto-iSCAT to track different energy carrier flow at various interfaces. Benefiting from the universality of scattering-based imaging, the transport of hot carriers, free carriers and thermal energy in gold, silicon and perovskites have been visualized. Then we mapped ultrafast dynamics of single nanoparticles and heterogeneous hot electron distributions in single 1D and 2D plasmonic resonators. The advantages in detection throughput and sensitivity help disclose new dynamical phenomena and the underlying mechanisms. Last but not least, we show that Femto-iSCAT enables efficient and versatile imaging of material function. We revealed distinct carrier and exciton decay processes through structure-dependent edge states in emerging optoelectronic semiconductors, establishing the fundamental relationship between microscopic structures and macroscopic photophysical properties.

We envision that more approaches for contrast enhancement in iSCAT will be adopted in Femto-iSCAT to further improve the transient signal level for more challenging targets. For instance, thin-film interference,[66] plasmon resonance[67] and photonic surfaces[68] are potentially applicable. In addition, Femto-iSCAT is sensitive to changes in the dielectric constant; thus, in principle, not only energy carriers but also molecules[69] or ions are possible to be detected. Furthermore, Femto-iSCAT is a promising tool for unraveling ultrafast energy and carrier transfer in single particles and even single molecules, which will promote researches on biosensing, optomechanics, heterogeneous catalysis and solar energy harvesting. Ultimately, we look forward to building a universal platform based on Femto-iSCAT to monitor ultrafast physical, chemical and biological processes on nanometer and femtosecond spatiotemporal scales.

## METHODS

**Sample preparation.** The monolayer graphene was purchased from ACS Material. The 10-nm thin gold film was fabricated by evaporation onto a cleaned glass coverslip. The prime-grade n-doped silicon wafer was purchased from Sibranch Microelectronics Technology Co. The $CsPbBr_3$ microcrystals were prepared by dissolving $CsPbBr_3$ powder (Polymer Light Technology Co.) at 10 mg/mL in N,N-dimethyl formamide (DMF, Sigma–Aldrich) followed by drop casting on a cleaned glass coverslip and crystallization. All gold nanoparticles were drop-cast on cleaned indium tin oxide (ITO)-coated glass slides (SPI Supplies). The 20 nm gold nanospheres were purchased from Ted Pella, Inc. The gold nanowires were synthesized using a three-step seeding approach.[70] The gold nanoplates were prepared using a thermal aqueous solution approach.[71] The layered 2D $WS_2$ crystal was mechanically exfoliated from a bulk crystal (2D Semiconductors) and then transferred onto a sapphire substrate. The mono- and few-layer flakes were verified by optical contrast and fluorescence spectroscopy.

**Experimental setup.** The output of a femtosecond Ti:sapphire amplifier (Solstice Ace, Spectra-Physics) was split into two beams, one providing the pump through an optical parametric amplifier (TOPAS Prime, Spectra-Physics) and the other supplying the probe. The pump beam was modulated at 10 Hz with a chopper (MC200B-EC, Thorlabs). The pump-probe time delays were controlled by an optical delay line (ODL300/M, Thorlabs). The pump beam was either focused or collimated onto the sample by an objective (S Plan Fluor ELWD 60x/0.7 NA, Nikon), depending on the pump mode. The probe beam was focused in the back focal plane of the objective to provide wide-field illumination. The pump and probe beams were spatially overlapped. Light scattered by the sample and reflected from the interface was collected by the same objective. The probe beam was spectrally and spatially filtered and detected by a scientific complementary metal oxide semiconductor camera (sCMOS; Dhyana 400D, Tucsen).

**Data analysis.** Setup automation and image acquisition were implemented with a custom-programmed software. The raw images were directly used to generate Femto-iSCAT images by a custom MATLAB script. Kinetic traces and signal profiles were extracted using a combination of ImageJ (Fiji) and MATLAB. Data analysis and plotting were performed using Origin 2016. Lifetime mappings were realized by a custom program combining Origin 2016 and JavaScript.

**ACKNOWLEDGEMENTS**

This work was mainly supported by the National Natural Science Foundation of China (22174064, 22034003, 21327902), Excellent Research Program of Nanjing University (ZYJH004) and State Key Laboratory of Analytical Chemistry for Life Science (5431ZZXM2002). We would like to thank Hua Li from the Prof. Weigao Xu group (Nanjing University) for providing the $WS_2$ samples.

**REFERENCES**

(1) Mehl, B. P.; Kirschbrown, J. R.; House, R. L.; Papanikolas, J. M. The End Is Different than the Middle: Spatially Dependent Dynamics in ZnO Rods Observed by Femtosecond Pump-Probe Microscopy. *J. Phys. Chem. Lett.* **2011**, *2*, 1777–1781.
(2) Snaider, J. M.; Guo, Z.; Wang, T.; Yang, M.; Yuan, L.; Zhu, K.; Huang, L. Ultrafast Imaging of Carrier Transport across Grain Boundaries in Hybrid Perovskite Thin Films. *ACS Energy Lett.* **2018**, *3*, 1402–1408.
(3) Wan, Y.; Guo, Z.; Zhu, T.; Yan, S.; Johnson, J.; Huang, L. Cooperative Singlet and Triplet Exciton Transport in Tetracene Crystals Visualized by Ultrafast Microscopy. *Nat. Chem.* **2015**, *7*, 785–792.
(4) Guo, Z.; Wan, Y.; Yang, M.; Snaider, J.; Zhu, K.; Huang, L. Long-Range Hot-Carrier Transport in Hybrid Perovskites Visualized by Ultrafast Microscopy. *Science* **2017**, *356*, 59–62.
(5) Zhu, Y.; Cheng, J. X. Transient Absorption Microscopy: Technological Innovations and Applications in Materials Science and Life Science. *J. Chem. Phys.* **2020**, *152*, No. 202901.
(6) Zhu, T.; Snaider, J. M.; Yuan, L.; Huang, L. Ultrafast Dynamic Microscopy of Carrier and Exciton Transport. *Annu. Rev. Phys. Chem.* **2019**, *70*, 219–244.
(7) Piland, G.; Grumstrup, E. M. High-Repetition Rate Broadband Pump-Probe Microscopy. *J. Phys. Chem. A* **2019**, *123*, 8709–8716.
(8) Schnedermann, C.; Sung, J.; Pandya, R.; Verma, S. D.; Chen, R. Y. S.; Gauriot, N.; Bretscher, H. M.; Kukura, P.; Rao, A. Ultrafast Tracking of Exciton and Charge Carrier Transport in Optoelectronic Materials on the Nanometer Scale. *J. Phys. Chem. Lett.* **2019**, *10*, 6727–6733.
(9) Bond, C.; Brown, D.; Freise, A.; Strain, K. A. *Interferometer Techniques for Gravitational-Wave Detection*; Springer International Publishing, 2016.
(10) Higuchi, M.; Wei, D.; Aketagawa, M. Displacement Measuring Interferometer for Sub-Nano Meter Resolution. *Meas. Sensors* **2021**, *18*, No. 100336.
(11) Park, Y. K.; Depeursinge, C.; Popescu, G. Quantitative Phase Imaging in Biomedicine. *Nat. Photonics* **2018**, *12*, 578–589.
(12) Lindfors, K.; Kalkbrenner, T.; Stoller, P.; Sandoghdar, V. Detection and Spectroscopy of Gold Nanoparticles Using Supercontinuum White Light Confocal Microscopy. *Phys. Rev. Lett.* **2004**, *93*, 3–6.
(13) Young, G.; Hundt, N.; Cole, D.; Fineberg, A.; Andrecka, J.; Tyler, A.; Olerinyova, A.; Ansari, A.; Marklund, E. G.; Collier, M. P.; Chandler, S. A.; Tkachenko, O.; Allen, J.; Crispin, M.; Billington, N.; Takagi, Y.; Sellers, J. R.; Eichmann, C.; Selenko, P.; Frey, L.; Riek, R.; Galpin, M. R.; Struwe, W. B.; Benesch, J. L. P.; Kukura, P. Quantitative Mass Imaging of Single Biological Macromolecules. *Science* **2018**, *360*, 423–427.

(14) Delor, M.; Weaver, H. L.; Yu, Q. Q.; Ginsberg, N. S. Imaging Material Functionality through Three-Dimensional Nanoscale Tracking of Energy Flow. *Nat. Mater.* **2020**, *19*, 56–62.

(15) Cole, D.; Young, G.; Weigel, A.; Sebesta, A.; Kukura, P. Label-Free Single-Molecule Imaging with Numerical-Aperture-Shaped Interferometric Scattering Microscopy. *ACS Photonics* **2017**, *4*, 211–216.

(16) Liebel, M.; Hugall, J. T.; Van Hulst, N. F. Ultrasensitive Label-Free Nanosensing and High-Speed Tracking of Single Proteins. *Nano Lett.* **2017**, *17*, 1277–1281.

(17) Cheng, C. Y.; Liao, Y. H.; Hsieh, C. L. High-Speed Imaging and Tracking of Very Small Single Nanoparticles by Contrast Enhanced Microscopy. *Nanoscale* **2019**, *11*, 568–577.

(18) Sun, J.; Melnikov, V. A.; Khan, J. I.; Mohammed, O. F. Real-Space Imaging of Carrier Dynamics of Materials Surfaces by Second-Generation Four-Dimensional Scanning Ultrafast Electron Microscopy. *J. Phys. Chem. Lett.* **2015**, *6*, 3884–3890.

(19) Delport, G.; Macpherson, S.; Stranks, S. D. Imaging Carrier Transport Properties in Halide Perovskites Using Time-Resolved Optical Microscopy. *Adv. Energy Mater.* **2020**, *10*, 1–13.

(20) Man, M. K. L.; Margiolakis, A.; Deckoff-Jones, S.; Harada, T.; Wong, E. L.; Krishna, M. B. M.; Madéo, J.; Winchester, A.; Lei, S.; Vajtai, R.; Ajayan, P. M.; Dani, K. M. Imaging the Motion of Electrons across Semiconductor Heterojunctions. *Nat. Nanotechnol.* **2017**, *12*, 36–40.

(21) Akselrod, G. M.; Deotare, P. B.; Thompson, N. J.; Lee, J.; Tisdale, W. A.; Baldo, M. A.; Menon, V. M.; Bulovic, V. Visualization of Exciton Transport in Ordered and Disordered Molecular Solids. *Nat. Commun.* **2014**, *5*, 1–8.

(22) Zhang, H.; Wei, J.; Zhang, X. G.; Zhang, Y. J.; Radjenovica, P. M.; Wu, D. Y.; Pan, F.; Tian, Z. Q.; Li, J. F. Plasmon-Induced Interfacial Hot-Electron Transfer Directly Probed by Raman Spectroscopy. *Chem* **2020**, *6*, 689–702.

(23) Guzelturk, B.; Lindenberg, A. M.; Utterback, J. K.; Coropceanu, I.; Kamysbayev, V.; Janke, E. M.; Zajac, M.; Yazdani, N.; Cotts, B. L.; Park, S.; Sood, A.; Lin, M. F.; Reid, A. H.; Kozina, M. E.; Shen, X.; Weathersby, S. P.; Wood, V.; Salleo, A.; Wang, X.; Talapin, D. V.; Ginsberg, N. S. Nonequilibrium Thermodynamics of Colloidal Gold Nanocrystals Monitored by Ultrafast Electron Diffraction and Optical Scattering Microscopy. *ACS Nano* **2020**, *14*, 4792–4804.

(24) Block, A.; Liebel, M.; Yu, R.; Spector, M.; Sivan, Y.; García De Abajo, F. J.; Van Hulst, N. F. Tracking Ultrafast Hot-Electron Diffusion in Space and Time by Ultrafast Thermomodulation Microscopy. *Sci. Adv.* **2019**, *5*, 1–8.

(25) Segovia, M.; Xu, X. High Accuracy Ultrafast Spatiotemporal Pump–Probe Measurement of Electrical Thermal Transport in Thin Film Gold. *Nano Lett.* **2021**, *21*, 7228–7235.

(26) Jacoboni, C.; Canali, C.; Ottaviani, G.; Alberigi Quaranta, A. A Review of Some Charge Transport Properties of Silicon. *Solid State Electron.* **1977**, *20*, 77–89.

(27) Jia, Y.; Wei, J.; Wang, K.; Cao, A.; Shu, Q.; Gui, X.; Zhu, Y.; Zhuang, D.; Zhang, G.; Ma, B.; Wang, L.; Liu, W.; Wang, Z.; Luo, J.; Wu, D. Nanotube-Silicon Heterojunction Solar Cells. *Adv. Mater.* **2008**, *20*, 4594–4598.

(28) Goodman, A. M. A Method for the Measurement of Short Minority Carrier Diffusion Lengths in Semiconductors. *J. Appl. Phys.* **1961**, *32*, 2550–2552.

(29) Najafi, E.; Ivanov, V.; Zewail, A.; Bernardi, M. Super-Diffusion of Excited Carriers in Semiconductors. *Nat. Commun.* **2017**, *8*, 1–7.

(30) Schröder, D. K. Carrier Lifetimes in Silicon. *IEEE Trans. Electron Devices* **1997**, *44*, 160–170.

(31) Xing, G.; Mathews, N.; Lim, S. S.; Lam, Y. M.; Mhaisalkar, S.; Sum, T. C. Long-Range Balanced Electron- and Hole-Transport Lengths in Organic-Inorganic CH3NH3PbI3. **2013**, *342*, 344–347.

(32) Alfurayj, I. A.; Li, Z.; Burda, C. Interfaces and Interfacial Carrier Dynamics in Perovskites. *J. Phys. Chem. C* **2021**, *125*, 15113–15124.

(33) Deng, S.; Snaider, J. M.; Gao, Y.; Shi, E.; Jin, L.; Schaller, R. D.; Dou, L.; Huang, L. Long-Lived Charge Separation in Two-Dimensional Ligand-Perovskite Heterostructures. *J. Chem. Phys.* **2020**, *152*, No. 044711.

(34) Deng, S.; Blach, D. D.; Jin, L.; Huang, L. Imaging Carrier Dynamics and Transport in Hybrid Perovskites with Transient Absorption Microscopy. *Adv. Energy Mater.* **2020**, *10*, 1–12.

(35) Guo, Z.; Manser, J. S.; Wan, Y.; Kamat, P. V.; Huang, L. Spatial and Temporal Imaging of Long-Range Charge Transport in Perovskite Thin Films by Ultrafast Microscopy. *Nat. Commun.* **2015**, *6*, 1–8.

(36) Huang, L.; Cheng, J. X. Nonlinear Optical Microscopy of Single Nanostructures. *Annu. Rev. Mater. Res.* **2013**, *43*, 213–236.

(37) Beane, G.; Devkota, T.; Brown, B. S.; Hartland, G. V. Ultrafast Measurements of the Dynamics of Single Nanostructures: A Review. *Reports Prog. Phys.* **2019**, *82*, No. 016401.

(38) Van Dijk, M. A.; Lippitz, M.; Orrit, M. Detection of Acoustic Oscillations of Single Gold Nanospheres by Time-Resolved Interferometry. *Phys. Rev. Lett.* **2005**, *95*, 1–4.

(39) Hartland, G. V. Measurements of the Material Properties of Metal Nanoparticles by Time-Resolved Spectroscopy. *Phys. Chem. Chem. Phys.* **2004**, *6*, 5263–5274.

(40) Liu, G.; Li, P.; Zhao, G.; Wang, X.; Kong, J.; Liu, H.; Zhang, H.; Chang, K.; Meng, X.; Kako, T.; Ye, J. Promoting Active Species Generation by Plasmon-Induced Hot-Electron Excitation for Efficient Electrocatalytic Oxygen Evolution. *J. Am. Chem. Soc.* **2016**, *138*, 9128–9136.

(41) Kim, M.; Lin, M.; Son, J.; Xu, H.; Nam, J. M. Hot-Electron-Mediated Photochemical Reactions: Principles, Recent Advances, and Challenges. *Adv. Opt. Mater.* **2017**, *5*, 1–21.

(42) Zou, N.; Chen, G.; Mao, X.; Shen, H.; Choudhary, E.; Zhou, X.; Chen, P. Imaging Catalytic Hotspots on Single Plasmonic Nanostructures via Correlated Super-Resolution and Electron Microscopy. *ACS Nano* **2018**, *12*, 5570–5579.

(43) Gargiulo, J.; Berté, R.; Li, Y.; Maier, S. A.; Cortés, E. From Optical to Chemical Hot Spots in Plasmonics. *Acc. Chem. Res.* **2019**, *52*, 2525–2535.

(44) Ditlbacher, H.; Hohenau, A.; Wagner, D.; Kreibig, U.; Rogers, M.; Hofer, F.; Aussenegg, F. R.; Krenn, J. R. Silver Nanowires as Surface Plasmon Resonators. *Phys. Rev. Lett.* **2005**, *95*, 1–4.

(45) Nicoletti, O.; De La Peña, F.; Leary, R. K.; Holland, D. J.; Ducati, C.; Midgley, P. A. Three-Dimensional Imaging of Localized Surface Plasmon Resonances of Metal Nanoparticles. *Nature* **2013**, *502*, 80–84.

(46) Viarbitskaya, S.; Teulle, A.; Marty, R.; Sharma, J.; Girard, C.; Arbouet, A.; Dujardin, E. Tailoring and Imaging the Plasmonic Local Density of States in Crystalline Nanoprisms. *Nat. Mater.* **2013**, *12*, 426–432.

(47) Li, L.; Cai, W.; Du, C.; Guan, Z.; Xiang, Y.; Ma, Z.; Wu, W.; Ren, M.; Zhang, X.; Tang, A.; Xu, J. Cathodoluminescence Nanoscopy of Open Single-Crystal Aluminum Plasmonic Nanocavities. *Nanoscale* **2018**, *10*, 22357–22361.

(48) Gao, Z.; Yin, L.; Fang, W.; Kong, Q.; Fan, C.; Kang, B.; Xu, J. J.; Chen, H. Y. Imaging Chladni Figure of Plasmonic Charge Density Wave in Real Space. *ACS Photonics* **2019**, *6*, 2685–2693.

(49) Merryweather, A. J.; Schnedermann, C.; Jacquet, Q.; Grey, C. P.; Rao, A. Operando Optical Tracking of Single-Particle Ion Dynamics in Batteries. *Nature* **2021**, *594*, 522–528.

(50) Park, J. S.; Lee, I. B.; Moon, H. M.; Ryu, J. S.; Kong, S. Y.; Hong, S. C.; Cho, M. Fluorescence-Combined Interferometric Scattering Imaging Reveals Nanoscale Dynamic Events of Single Nascent Adhesions in Living Cells. *J. Phys. Chem. Lett.* **2020**, *11*, 10233–10241.

(51) Hartland, G. V. Optical Studies of Dynamics in Noble Metal Nanostructures. *Chem. Rev.* **2011**, *111*, 3858–3887.

(52) Zhu, T.; Yuan, L.; Zhao, Y.; Zhou, M.; Wan, Y.; Mei, J.; Huang, L. Highly Mobile Charge-Transfer Excitons in Two-Dimensional WS2/Tetracene Heterostructures. *Sci. Adv.* **2018**, *4*, 1–9.

(53) Unuchek, D.; Ciarrocchi, A.; Avsar, A.; Watanabe, K.; Taniguchi, T.; Kis, A. Room-Temperature Electrical Control of Exciton Flux in a van Der Waals Heterostructure. *Nature* **2018**, *560*, 340–344.

(54) Ni, Z.; Bao, C.; Liu, Y.; Jiang, Q.; Wu, W. Q.; Chen, S.; Dai, X.; Chen, B.; Hartweg, B.; Yu, Z; Holman, Z.; Huang, J. Resolving spatial and energetic distributions of trap states in metal halide perovskite solar cells. *Science* **2020**, *367*, 1352–1358.

(55) Blancon, J. C.; Tsai, H.; Nie, W.; Stoumpos, C. C.; Pedesseau, L.; Katan, C.; Kepenekian, M.; Soe, C. M. M.; Appavoo, K.; Sfeir, M. Y.; Tretiak, S.; Ajayan, P. M.; Kanatzidis, M. G.; Even, J.; Crochet, J. J.; Mohite, A. D. Extremely Efficient Internal Exciton Dissociation through Edge States in Layered 2D Perovskites. *Science* **2017**, *355*, 1288–1292.

(56) Yang, B.; Hong, F.; Chen, J.; Tang, Y.; Yang, L.; Sang, Y.; Xia, X.; Guo, J.; He, H.; Yang, S.; Deng, W.; Han, K. Colloidal Synthesis and Charge-Carrier Dynamics of Cs2AgSb1−yBiyX6 (X: Br, Cl; 0 ≤y ≤1) Double Perovskite Nanocrystals. *Angew. Chem. Int. Ed.* **2019**, *58*, 2278–2283.

(57) Guo, Z.; Wu, X.; Zhu, T.; Zhu, X.; Huang, L. Electron-Phonon Scattering in Atomically Thin 2D Perovskites. *ACS Nano* **2016**, *10*, 9992–9998.

(58) Jin, H.; Debroye, E.; Keshavarz, M.; Scheblykin, I. G.; Roeffaers, M. B. J.; Hofkens, J.; Steele, J. A. It's a Trap! On the Nature of Localised States and Charge Trapping in Lead Halide Perovskites. *Mater. Horizons* **2020**, *7*, 397–410.

(59) Kang, J.; Wang, L. W. High Defect Tolerance in Lead Halide Perovskite CsPbBr3. *J. Phys. Chem. Lett.* **2017**, *8*, 489–493.

(60) Cunningham, P. D.; McCreary, K. M.; Jonker, B. T. Auger Recombination in Chemical Vapor Deposition-Grown Monolayer WS2. *J. Phys. Chem. Lett.* **2016**, *7*, 5242–5246.

(61) Cunningham, P. D.; Hanbicki, A. T.; McCreary, K. M.; Jonker, B. T. Photoinduced Bandgap Renormalization and Exciton Binding Energy Reduction in WS2. *ACS Nano* **2017**, *11*, 12601–12608.

(62) Wang, H.; Zhang, C.; Rana, F. Surface Recombination Limited Lifetimes of Photoexcited Carriers in Few-Layer Transition Metal Dichalcogenide MoS2. *Nano Lett.* **2015**, *15*, 8204–8210.

(63) Ruppert, C.; Chernikov, A.; Hill, H. M.; Rigosi, A. F.; Heinz, T. F. The Role of Electronic and Phononic Excitation in the Optical Response of Monolayer WS2 after Ultrafast Excitation. *Nano Lett.* **2017**, *17*, 644–651.

(64) Eroglu, Z. E.; Comegys, O.; Quintanar, L. S.; Azam, N.; Elafandi, S.; Mahjouri-Samani, M.; Boulesbaa, A. Ultrafast Dynamics of Exciton Formation and Decay in Two-Dimensional Tungsten Disulfide (2D-WS2) Monolayers. *Phys. Chem. Chem. Phys.* **2020**, *22*, 17385–17393.

(65) Yuan, L.; Huang, L. Exciton Dynamics and Annihilation in WS2 2D Semiconductors. *Nanoscale* **2015**, *7*, 7402–7408.

(66) Sevenler, D.; Daaboul, G. G.; Ekiz Kanik, F.; Ünlü, N. L.; Ünlü, M. S. Digital Microarrays: Single-Molecule Readout with Interferometric Detection of Plasmonic Nanorod Labels. *ACS Nano* **2018**, *12*, 5880–5887.

(67) Shi, Z.; Huang, J.; Huang, X.; Huang, Y.; Wu, L.; Li, Q. Resonant Scattering Enhanced Interferometric Scattering Microscopy. *Nanoscale* **2020**, *12*, 7969–7975.

(68) Li, N.; Canady, T. D.; Huang, Q.; Wang, X.; Fried, G. A.; Cunningham, B. T. Photonic Resonator Interferometric Scattering Microscopy. *Nat. Commun.* **2021**, *12*, 1–9.

(69) Celebrano, M.; Kukura, P.; Renn, A.; Sandoghdar, V. Single-Molecule Imaging by Optical Absorption. *Nat. Photonics* **2011**, *5*, 95–98.

(70) Kim, F.; Sohn, K.; Wu, J.; Huang, J. Chemical Synthesis of Gold Nanowires in Acidic Solutions. *J. Am. Chem. Soc.* **2008**, *130*, 14442–14443.

(71) Huang, W. L.; Chen, C. H.; Huang, M. H. Investigation of the Growth Process of Gold Nanoplates Formed by Thermal Aqueous Solution Approach and the Synthesis of Ultra-Small Gold Nanoplates. *J. Phys. Chem. C* **2007**, *111*, 2533–2538.

**FIGURES**

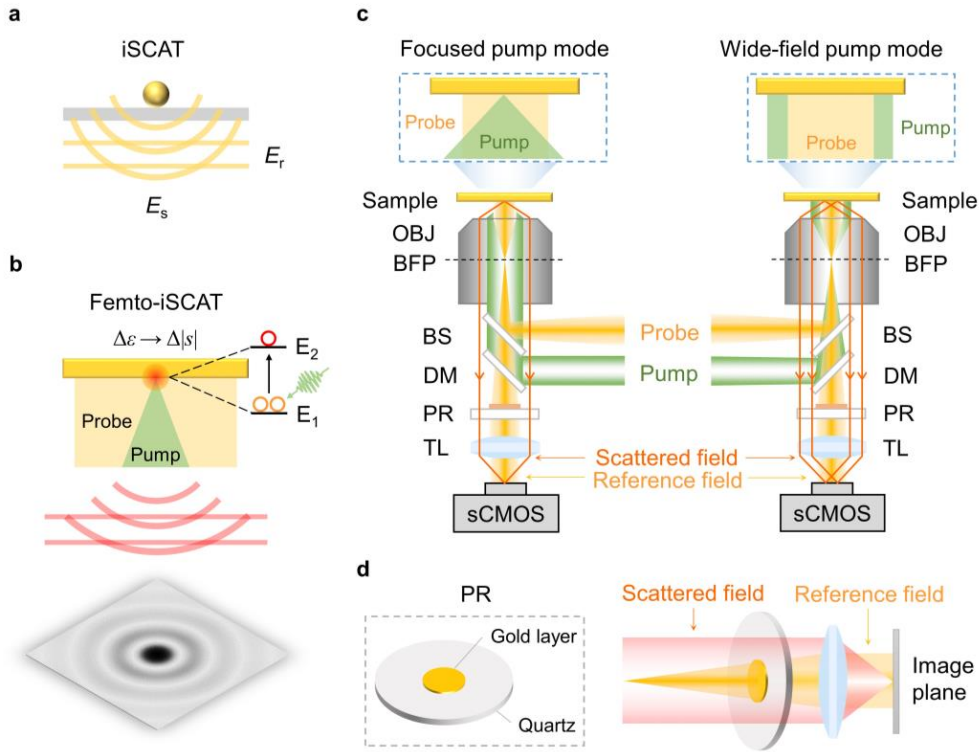

**Figure 1.** Principle and setup of Femto-iSCAT. (a) Schematic of iSCAT. $E_s$, the scattered field of the sample; $E_r$, the reference field from the interface. (b) Schematic of Femto-iSCAT. A femtosecond pump pulse excites the sample ($E_1 \rightarrow E_2$) thus changes the local dielectric constants ($\Delta\varepsilon$), leading to the changes to the scattering amplitude ($\Delta|s|$), which is detected by a wide-field probe pulse. The PSF is a 2D Bessel function. (c) Simplified sketch of the experimental setup. Two implementations are provided: focused pump mode and wide-field pump mode. OBJ, objective; BFP, back focal plane; BS, 50:50 beam splitter; DM, dichroic mirror; TL, tube lens; PR, partial reflector; sCMOS, scientific complementary metal oxide semiconductor camera. The scattered field in wide-field pump mode originates from superposition of scattering sources, two of which are depicted as representatives. (d) Schematic of the structure and function of the PR.

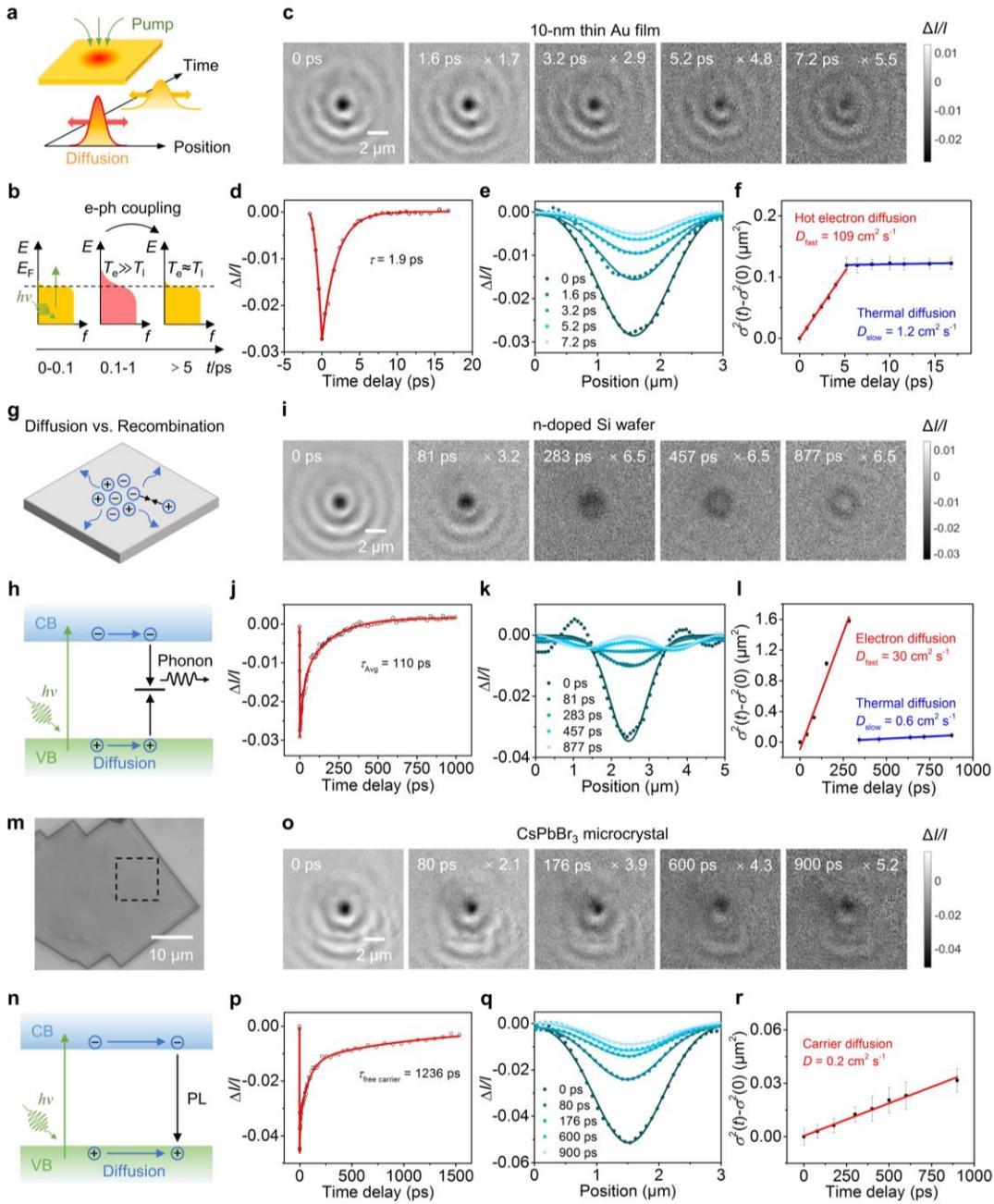

**Figure 2.** Carrier or heat transport in 10-nm thin Au films, n-doped Si wafers and CsPbBr$_3$ microcrystals. (a) Schematic of carrier or heat transport. (b) Illustration of hot electron dynamics in Au films. $E_F$, Fermi level; f, occupation of electrons. (c-f, i-l, o-r) Femto-iSCAT time series, transient dynamics, spatial profiles with Gaussian fittings and the variances evolution of the profiles of (c-f) the 10-nm thin Au film, (i-l) the n-doped Si wafer and (o-r) the CsPbBr$_3$ microcrystal. (g) Schematic of carrier diffusion and recombination in semiconductors. (h, n) Illustration of carrier dynamics in (h) n-doped Si wafers and (n) CsPbBr$_3$ microcrystals. VB, valence bands; CB, conduction bands. (m) Optical image of a CsPbBr$_3$ microcrystal. The black dashed square is the region of interest in (o).

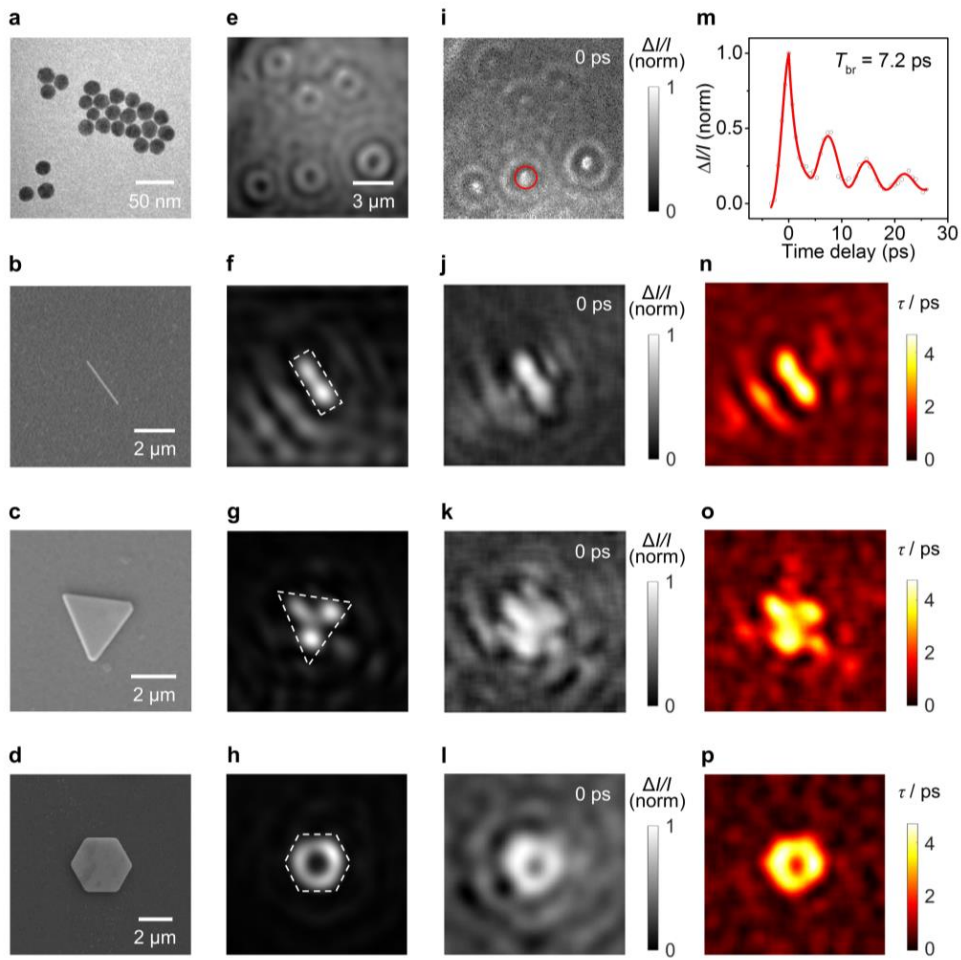

**Figure 3.** Hot electron dynamics and acoustic vibrations in single plasmonic resonators. (a) Transmission electron microscopy (TEM) image of 20 nm gold nanospheres. (b-d) Scanning electron microscopy (SEM) images of (b) a single gold nanowire, (c) a single triangular gold nanoplate and (d) a single hexagonal gold nanoplate. (e-h) iSCAT images of the corresponding particles in (a-d). The white dashed lines are the outlines of the particles. (i-l) Femto-iSCAT images of the corresponding particles in (e-h) at 0 ps, which denote the hot electron distributions. (m) Transient dynamics of a single gold nanosphere (marked by a solid red circle in (i)). (n, o, p) Hot electron lifetime ($\tau$) maps of the corresponding particles in (f-h). Pump fluences are 50 μJ cm$^{-2}$.

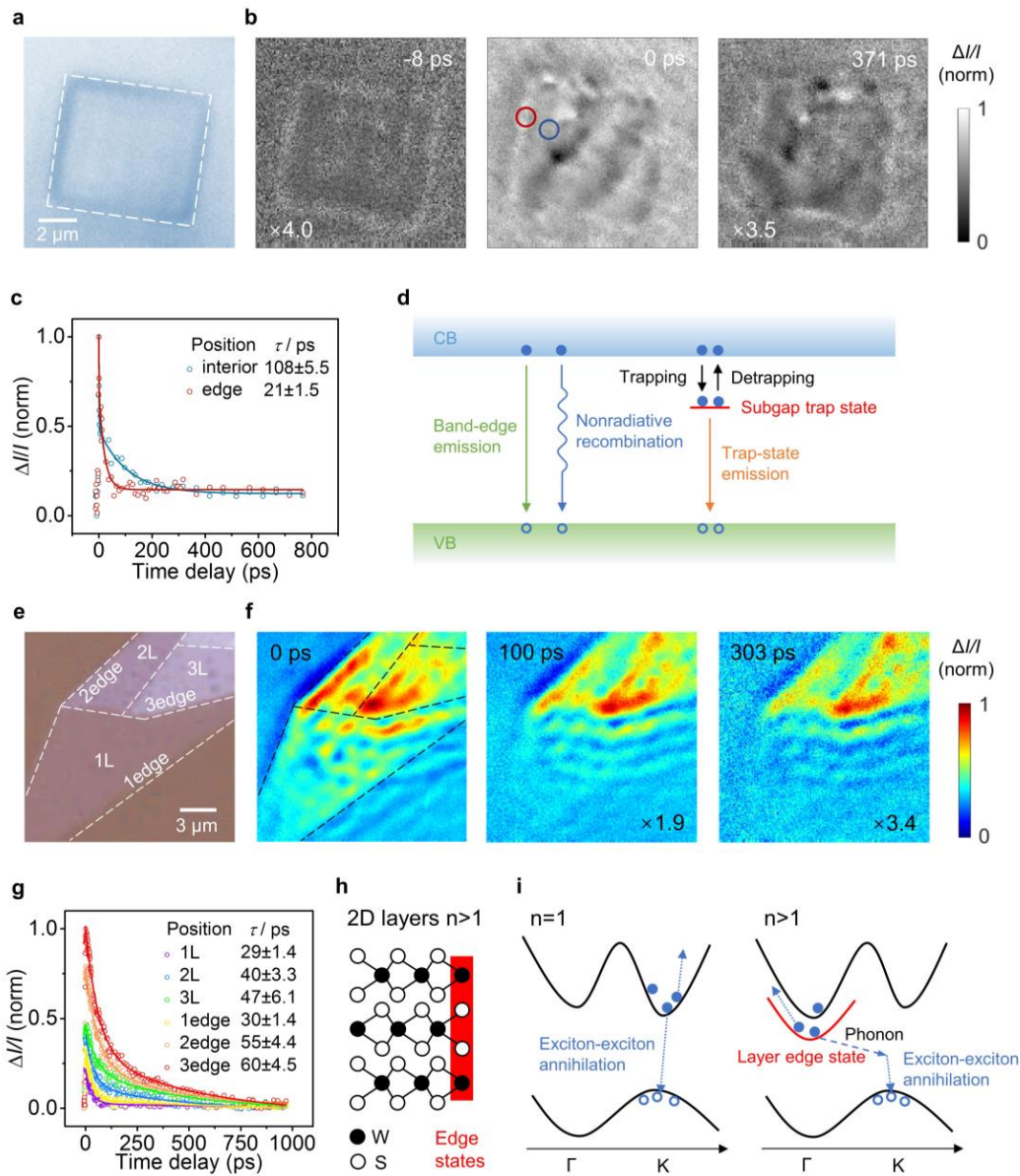

**Figure 4.** Edge state dynamics in CsPbBr$_3$ microcrystals and 2D WS$_2$ layers. (a) False-color optical image of a CsPbBr$_3$ microcrystal. (b) Femto-iSCAT time series of the CsPbBr$_3$ microcrystal. Pump fluence is 5 µJ cm$^{-2}$ (Pump: 480 nm). (c) Transient dynamics at the interior and the edge (the blue and red circles in (b)) of the CsPbBr$_3$ microcrystal. (d) Schematic of various carrier decay pathways in CsPbBr$_3$ microcrystals. (e) False-color optical image of an exfoliated WS$_2$ crystal. (f) Femto-iSCAT time series of the WS$_2$ crystal. Pump fluence is 10 µJ cm$^{-2}$ (Pump: 532 nm). (g) Transient dynamics of different positions on the WS$_2$ crystal. (h) Schematic of edge states in 2D layers for n>1 (n denotes layer number). (i) Schematic of nonradiative exciton recombination in monolayer and few-layer WS$_2$.